\documentclass[conference]{IEEEtran}
\IEEEoverridecommandlockouts

\usepackage{cite}
\usepackage{array}
\usepackage{graphicx}
\usepackage{amsmath,amssymb,bm,amsthm,amsfonts}
\usepackage{subfig}
\usepackage[linesnumbered,ruled,vlined]{algorithm2e}
\usepackage{gensymb}
\usepackage{multirow}
\usepackage{diagbox}

\begin{document}

\title{Cheating-Resilient Incentive Scheme for Mobile Crowdsensing Systems}

\author{\IEEEauthorblockN{Cong Zhao\IEEEauthorrefmark{1}, Xinyu Yang\IEEEauthorrefmark{1}, Wei Yu\IEEEauthorrefmark{2}, Xianghua Yao\IEEEauthorrefmark{1}, Jie Lin\IEEEauthorrefmark{1} and Xin Li\IEEEauthorrefmark{1}}
\IEEEauthorblockA{\IEEEauthorrefmark{1}Xi'an Jiaotong University}
\IEEEauthorblockA{\IEEEauthorrefmark{2}Towson University}}

\maketitle

\begin{abstract}

Mobile Crowdsensing is a promising paradigm for ubiquitous sensing, which explores the tremendous data collected by mobile smart devices with prominent spatial-temporal coverage. As a fundamental property of Mobile Crowdsensing Systems, temporally recruited mobile users can provide agile, fine-grained, and economical sensing labors, however their self-interest cannot guarantee the quality of the sensing data, even when there is a fair return. Therefore, a mechanism is required for the system server to recruit well-behaving users for credible sensing, and to stimulate and reward more contributive users based on sensing truth discovery to further increase credible reporting. In this paper, we develop a novel Cheating-Resilient Incentive (CRI) scheme for Mobile Crowdsensing Systems, which achieves credibility-driven user recruitment and payback maximization for honest users with quality data. Via theoretical analysis, we demonstrate the correctness of our design. The performance of our scheme is evaluated based on extensive real-world trace-driven simulations. Our evaluation results show that our scheme is proven to be effective in terms of both guaranteeing sensing accuracy and resisting potential cheating behaviors, as demonstrated in practical scenarios, as well as those that are intentionally harsher.

\end{abstract}


\IEEEpeerreviewmaketitle

\section{Introduction} \label{sec:introduction}

The proliferation of mobile smart devices has promoted the development of Mobile Crowdsensing Systems (MCSs), a promising paradigm for agile, fine-grained, and economical sensing with prominent spatial-temporal coverage \cite{Lane2015A}. Exploring people-centric data collected by smart devices with enriched sensors (\emph{e.g.} Global Positioning System, gyroscope and microphone), a growing number of MCS prototypes have been developed to support applications, including urban sensing \cite{Gao2014jigsaw}, environmental monitoring \cite{Capezzuto2014A} and mobile social networking \cite{bakht2012searchlight}.

Observing that the data source of MCSs is a set of personal mobile devices temporally recruited, the self-interested nature of mobile users needs to be taken into account for MCS implementations. From the perspective of mobile users considering the potential costs (\emph{e.g.} physical labor, device battery life, and network bandage usage), participation is unlikely in an MCS sensing task unless there was a considerable payback under the rational person hypothesis. From the perspective of the MCS server, the credibility of reported observations from temporally recruited users is not guaranteed (\emph{i.e.} users may cheat in sensing tasks just for the payback without reporting quality data, which is referred to as Cheating Behavior in this paper), even when it pays fairly for data acquisition. Intuitively, we can see that: (i) the MCS server needs to recruit credible users in sensing tasks, and (ii) honest responses deserve substantial reward while dishonest reporting requires reprimand. A mechanism satisfying these two requirements simultaneously is necessary for MCS implementations.

User Incentive schemes \cite{Gao2015A} aim at encouraging self-interested users to participate in system tasks by rewarding monetary or tradable paybacks. Generally speaking, existing schemes in participatory systems model the user incentive process as an optimization problem for either the system server \cite{Lee2010Sell,Wang2013An} or users \cite{Luo2014Fairness,Sun2014Heterogeneous} by designing mechanisms based on auction or game theory. Nonetheless, research on incentive mechanisms that considers active cheating behaviors from self-interested users is relatively limited. Schemes in \cite{Zhao2014How} and \cite{Zhang2014Free} aim to guarantee the `bidder's truthfulness' in designed auctions. Also, incentive schemes were proposed to evaluate the quality of user reports \cite{Gao2013On,Faltings2014Incentive}. However, these schemes cannot be directly deployed in highly dynamic and opportunistic MCSs. To stimulate the service time of MCS participators, a Stackelberg game-based incentive mechanism is proposed in \cite{Yang2012Crowdsourcing}, which maximizes the utility of the MCS platform and proves that a best strategy for all self-interested participators can be centrally determined. However, since the purpose of this work is to stimulate user participation, and no sensing data quality related factor is considered, it cannot solve the dishonest user reporting issue. An incentive mechanism that encourages quality data reporting is necessary to guarantee the usability of MCSs.

In this paper, we develop a novel Cheating-Resilient Incentive (CRI) scheme for MCSs, which guarantees the accuracy of crowdsensing tasks while encouraging mobile users to provide quality data without cheating for maximum paybacks. Our contributions are summarized as follows:

\begin{itemize}

\item Based on the participation-driven incentivization in \cite{Yang2012Crowdsourcing}, we develop a reputation-driven method for the MCS server to recruit the most credible users autonomously according to their historical behaviors. Meanwhile, recruited users can obtain maximum paybacks only when they contribute no less than expected. We demonstrate the correctness of our design with theoretical analysis.

\item We introduce the truth discovery technique \cite{Miao2015Cloud} into the user incentive issue to evaluate the actual contribution of recruited users in MCS tasks. The adaptive truth discovery guarantees the accuracy of crowdsensing while providing a baseline for user contribution evaluation.

\item Through extensive trace-driven simulations, we evaluate the performance of CRI. The simulation results validate the effectiveness of CRI with respect to both quality-driven user stimulation and cheating behavior resistance.

\end{itemize}

The remainder of the paper is organized as follows: In Section~\ref{sec:modeling}, we present the MCS model and the description of the user incentive problem. In Section~\ref{sec:scheme}, we present the CRI scheme in detail. In Section~\ref{sec:evaluation}, we present the evaluation results of our designed scheme via extensive trace-driven simulations. Finally, we conclude this paper in Section~\ref{sec:conclusion}.

\begin{figure}
\centering
\includegraphics[width=2.5in]{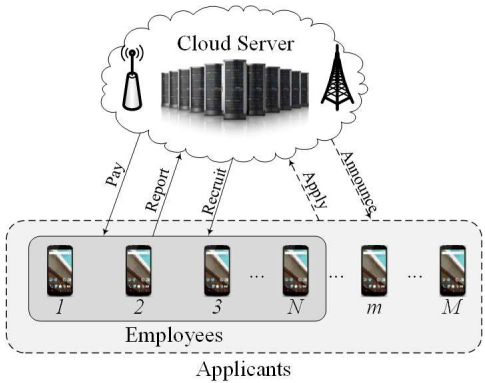}
\caption{System Architecture} \label{fig:systemarc}
\vspace{-1em}
\end{figure}

\section{System Model and Problem Definition} \label{sec:modeling}

\subsection{MCS Architecture}\label{subsec:mcsarchitecture}

As shown in Figure~\ref{fig:systemarc}, a general MCS consists of a cloud server $s$ and a set of registered mobile users $\mathcal{P}=\{1,2,\ldots,M\}$, where $M \geqslant 2$. Any user $i\in\mathcal{P}$ can communicate with $s$ via either cellular or WiFi access points.

For a specific sensing target, $s$ can announce a task $\tau$ to all users in $\mathcal{P}$. Receiving the announcement, any user $i\in\mathcal{P}$ who is interested in $\tau$ can reply with its reputation $r_i$ as an application, which reflects its behavior in historical tasks. According to all applications received, $s$ can determine the total reward $R$ and the set of applicants $\mathcal{E}=\{1,2,\ldots,N\}\subseteq\mathcal{P}$ to be the final employees of $\tau$. All employees $\in\mathcal{E}$ then conduct sensing obligations required by $\tau$ and report their observations $\mathcal{O}=\{o_1,o_2,\ldots,o_N\}$ to $s$. Based on all reports received, $s$ can discover $\tau$'s sensing truth $o_{\tau}$, and evaluate the contribution $\mathcal{C}=\{c_1,c_2,\ldots,c_N\}$ of all employees separately, which determines their paybacks $\mathcal{G}=\{g_1,g_2,\ldots,g_N\}$ for $\tau$ and corresponding reputation adjustments.

\subsection{User Incentive Problem} \label{subsec:problem}

The purpose of incentive mechanisms in MCSs is to stimulate mobile user activity through paybacks for participating in sensing tasks. As the decision maker, $s$ prefers to maximize paybacks of the most contributive users as stimulation, while minimizing the effect of potential cheating behaviors (\emph{e.g.} intentionally reporting random data just for the payback), to achieve accurate sensing.

To guarantee the quality of sensing reports, $s$ needs to consider two issues: (i) the selection of employees for a given task should be based on the applicant's reputation, and (ii) the reputation adjustment and payback to an employee should be based on their contribution in the current task. Therefore, our user incentive scheme should solve the following problems:
\begin{enumerate}
\item How to define and manage the user reputation?
\item How to recruit employees based on their reputations?
\item How to evaluate an employee's contribution?
\item How to quantify and maximize an employee's payback according to its contribution?
\end{enumerate}

\section{The Cheating-Resilient Incentive Scheme} \label{sec:scheme}

In this section, we develop the Cheating-Resilient Incentive (CRI) scheme to address the four problems outlined in Subsection~\ref{subsec:problem}. Before the construction of CRI, we first provide the formal definitions of both \textit{user reputation} and \textit{employee payback}. Then, we present a mechanism for reputation-driven employee recruitment. After that, we address the issue of how to evaluate an employee's contribution and how to quantify and maximize an employee's payback according to its contribution.

\subsection{Definitions}

In this paper, we formally define user reputation and employee payback as follows.

\textbf{User Reputation:} Intuitively, we treat the reputation as a user's credibility for offering quality reports. In fact, once recruited, the user is responsible for providing an actual contribution to the sensing task in proportion to its reputation.

For any mobile user $i\in \mathcal{P}$, the reputation of $i$, denoted as $r_i$ ($0\leqslant r_i\leqslant 1$), will be adjusted by $s$ whenever $i$ participates in a sensing task $\tau$ by
\begin{equation}\label{eq:reputation}
r_i=\alpha r'_i+(1-\alpha)\frac{c_i}{\bar{c}_i},
\end{equation}
where $r'_i$ denotes the reputation of $i$ before it participates in $\tau$, $\bar{c}_i$ denotes $i$'s expected contribution to $\tau$ (estimated by $s$ considering $r'_i$, discussed latter), $c_i$ denotes $i$'s actual contribution to $\tau$ (evaluated by $s$, discussed latter), and $0\leqslant\alpha\leqslant1$ determines the sensitivity of $i$'s reputation adjustment. After the user's registration, $s$ issues any $i\in \mathcal{P}$ a value $r_0$ as its initial reputation.

\textbf{Employee Payback:} Intuitively, it is natural to determine an employee's payback according to its contribution. Also, considering that the reputation of an employee only denotes the probability of its good behavior, we take the potential quality risk into account for the payback determination.

Inspired by the sensing time-driven model in \cite{Yang2012Crowdsourcing}, for an employee $i\in\mathcal{E}$ of task $\tau$, we define its reputation-driven payback $g_i$ as
\begin{equation}\label{eq:payback}
g_i=\frac{c_i}{\sum_{j\in\mathcal{E}}c_j}R-\frac{1-r_i}{r_i}c_i,
\end{equation}
where $R>0$ denotes the total reward of $\tau$. For a fixed $R$, $g_i$ is subjected to both the ratio of $i$'s contribution, $\frac{c_i}{\sum_{j\in\mathcal{E}}c_j}R$, and the potential quality risk for recruiting $i$, $\frac{1-r_i}{r_i}c_i$.

Based on the definitions above, we develop CRI, which consists of three components: (i) \textit{Employee Recruitment}, (ii) \textit{Contribution Evaluation}, and (iii) \textit{Payback Determination}.

\subsection{Employee Recruitment}
\label{subsec:employeerecruitment}

\begin{algorithm}[!t]
\caption{Reputation-Driven Employee Recruitment}
\label{alg:recruit}
\KwIn{\\
$\mathcal{R}$: reputations of $\forall i\in\mathcal{P}$\;}
\KwOut{\\
$\mathcal{E}$: recruited employees\;
$\bar{\mathcal{C}}$: expected contributions of $\forall i\in\mathcal{E}$.}
\smallskip
\hrule
\smallskip
put $\forall r_i\in\mathcal{R}$ in descending order\;
$\bar{\mathcal{C}}\gets\varnothing$, $\mathcal{E}\gets\{1,2\}$, $i\gets3$\;
\While{$i\leqslant\lvert\mathcal{P}\rvert$ \&\& $r_i>\frac{\lvert\mathcal{E}\rvert-1}{\sum_{j\in\mathcal{E}}\frac{1-r_j}{r_j}+\lvert\mathcal{E}\rvert-1}$}
{
$\mathcal{E}\gets\mathcal{E}\cup\{i\}$\;
$i^{++}$\;
}
\For{$\forall i\in\mathcal{P}$}
{
\If{$i\in\mathcal{E}$}{$\bar{c}_i=\frac{(\lvert\mathcal{E}\rvert-1)R}{\sum_{j\in\mathcal{E}}\frac{1-r_j}{r_j}}(1-\frac{(\lvert\mathcal{E}\rvert-1)\frac{1-r_i}{r_i}}{\sum_{j\in\mathcal{E}}\frac{1-r_j}{r_j}})$\;}
\Else{$\bar{c}_i=0$\;}
$\bar{\mathcal{C}}\gets\bar{\mathcal{C}}\cup\bar{c}_i$\;
}
\Return{$\mathcal{E}$, $\bar{\mathcal{C}}$}\;
\end{algorithm}

We now address the issue of how to recruit employees based on their reputations. After the announcement of task $\tau$, receiving all applications $\mathcal{R}=\{r_1,r_2,\ldots,r_M\}$ (see Subsection~ \ref{subsec:mcsarchitecture}), $s$ prefers to recruit employees with top reputations. Nonetheless, according to Equation~(\ref{eq:payback}), it is possible that an applicant could obtain no payback if the total reward budget is low. Therefore, $s$ needs to determine the final employees considering their expected paybacks $\bar{\mathcal{G}}=\{\bar{g}_1,\bar{g}_2,\ldots,\bar{g}_M\}$, which are determined by their expected contributions $\bar{\mathcal{C}}=\{\bar{c}_1,\bar{c}_2,\ldots,\bar{c}_M\}$.

For effective stimulations, it is necessary for $s$ to maximize the expected payback that a well-behaving employee can have. According to Equation~(\ref{eq:payback}), it is easy to get that $g_i$ is second-order continuous differentiable on $\bar{c}_i$, and the second derivative of $g_i$ on $\bar{c}_i$ is:

\begin{equation} \label{eq:deri2}
\ddot{g}_i(\bar{c}_i)=-\frac{2R\sum_{j\in \mathcal{P}\backslash\{i\}}\bar{c}_j}{(\sum_{j\in \mathcal{P}}\bar{c}_j)^3}.
\end{equation}

When $\bar{c}_i\geqslant0$, there is $\ddot{g}_i(\bar{c}_i)<0$, and $g_i(\bar{c}_i)$ is a concave function. Therefore, $g_i(\bar{c}_i)$ has a unique maximum value when $\vert \mathcal{P}\vert\geqslant2$. We call such a value $\bar{g}_i$ as $i$'s expected payback, which can be calculated whenever $\dot{g}_i(\bar{c}_i)=0, \bar{c}_i>0$ has a solution.

Therefore, from the perspective of $s$, any $i\in\mathcal{P}$ that is finally recruited should satisfy following restrictions:

\begin{equation} \label{eq:recruitrestrictions}
\begin{cases}
\bar{c}_i=\sqrt{\frac{r_iR\sum_{j\in \mathcal{E}\backslash\{i\}}\bar{c}_j}{1-r_i}}-\sum_{j\in \mathcal{E}\backslash\{i\}}\bar{c}_j>0;\\
\bar{g}_i=\frac{\bar{c}_i}{\sum_{j\in\mathcal{E}}\bar{c}_j}R-\frac{1-r_i}{r_i}\bar{c}_i>0.
\end{cases}
\end{equation}

To allow $s$ to autonomously recruit as many credible users as possible with a fixed total reward, we develop Algorithm~\ref{alg:recruit} based on the NE computation algorithm in the STD game \cite{Yang2012Crowdsourcing} to determine final employees just based on their reputations. Here, all employees recruited will have the maximum payback only if they contribute as expected.

According to \cite{Yang2012Crowdsourcing}, we have Propositions 1 and 2 for Algorithm~\ref{alg:recruit} as follows.

\vspace{0.1em}
\textbf{Proposition 1.} \textit{Any $i\in\mathcal{P}$ that is not recruited by Algorithm \ref{alg:recruit} gets the maximum payback by not participating in $\tau$.}
\vspace{0.1em}

\vspace{0.1em}
\textbf{Proposition 2.} \textit{Any $i\in\mathcal{E}$ that is recruited by Algorithm \ref{alg:recruit} can obtain the maximum payback only if it contributes as expected in $\tau$.}
\vspace{0.1em}

Because of the page limitation, please refer to Theorem 1 and 2 in \cite{Yang2012Crowdsourcing} for specific proofs.

Until now, $s$ can recruit the most credible employees $\mathcal{E}$ and compute their expected contributions $\bar{\mathcal{C}}$ only based on all received applications $\mathcal{R}$, and $\bar{\mathcal{C}}$ is treated as one of the metrics for the payback determination.

\subsection{Contribution Evaluation}
\label{subsec:contributionevaluation}

In the following, we address the issue of how to evaluate an employee's contribution. After employee recruitment, $s$ announces a detailed task description of $\tau$ to all $i\in\mathcal{E}$. The employees then reply corresponding reports containing their observations $\mathcal{O}=\{o_1,o_2,\ldots,o_N\}$\footnote{We treat each employee's task observation as a single-dimension value for concision, which can be expanded in different sensing scenarios.}. Based on $\mathcal{O}$, $s$ discovers the sensing truth $o_{\tau}$, and evaluates the actual contributions $\mathcal{C}=\{c_1,c_2,\ldots,c_N\}$ of all $i\in\mathcal{E}$.

Because there is no ground truth in our MCS scenario, $s$ needs to discover the sensing truth $o_{\tau}$, based on $\mathcal{O}$, as the baseline to evaluate employees' actual contributions. Considering the potential conflict in reported observations and the difference in employee reputations, we develop Algorithm~\ref{alg:sensingtruthdiscovery} based on the general truth discovery framework in \cite{Miao2015Cloud} to allow $s$ to compute $o_{\tau}$ and to evaluate actual employee contributions $\mathcal{C}$ at the same time.

According to Algorithm~\ref{alg:sensingtruthdiscovery}, each actual contribution $c_i\in\mathcal{C}$ is subjected to both the distance between $o_i$ and $o_{\tau}$, and $i$'s reputation $r_i$. We treat observations from employees with higher reputations as more credible, and an employee's contribution will be higher if its observation is closer to $o_{\tau}$. $\mathcal{C}$ is treated as another metric for the payback determination.

\begin{algorithm}[htp]
\caption{Weighted Sensing Truth Discovery}
\label{alg:sensingtruthdiscovery}
\KwIn{\\
$\mathcal{O}$: observations of $\forall i\in\mathcal{E}$\;
$\mathcal{R_{\mathcal{E}}}$: reputations of $\forall i\in\mathcal{E}$\;
$\epsilon$: convergence threshold\;}
\KwOut{\\
$o_{\tau}$: discovered sensing truth\;
$\mathcal{C}$: actual contributions of $\forall i\in\mathcal{E}$.}
\smallskip
\hrule
\smallskip
compute the standard deviation of $\mathcal{O}$: $std_{\mathcal{O}}$\;
initialize $o_{\tau}$ as a random value\;
initialize $\forall c_i\in\mathcal{C}$ as $0$\;
\Repeat{
$\lvert o_\tau-o'_\tau \rvert<\epsilon$}
{\For{$\forall i\in\mathcal{E}$}{
$c_i=\log(\displaystyle{\frac{\sum_{j\in\mathcal{E}}\frac{(o_j-o_\tau)^2}{std_\mathcal{O}r_j}}{\frac{(o_i-o_\tau)^2}{std_\mathcal{O}r_i}}})$\;
}
$o'_\tau=o_\tau$\;
$o_\tau=\displaystyle{\frac{\sum_{j\in\mathcal{E}}c_jo_j}{\sum_{j\in\mathcal{E}}c_j}}$\;}
\For{$\forall c_i\in\mathcal{C}$}{$c_i=\displaystyle{\frac{c_i}{\sum_{c_j\in\mathcal{C}}c_j}}$\;}
\Return{$o_{\tau}$, $\mathcal{C}$}\;
\end{algorithm}

\subsection{Payback Determination}
\label{subsec:paybackdetermination}

We now answer the question about how to quantify and maximize an employee's payback according to its contribution. Based on the expected contributions $\bar{\mathcal{C}}$ and the actual contributions $\mathcal{C}$, $s$ can determine the final paybacks $\mathcal{G}=\{g_1,g_2,\ldots,g_N\}$ for all $i\in\mathcal{E}$, and update employees' reputations $\mathcal{R_{\mathcal{E}}}$ according to their behaviors in $\tau$.

For all $i\in\mathcal{E}$, to compare $\bar{c}_i$ and $c_i$, we set the total reward $R=\frac{\sum_{j\in\mathcal{E}}\frac{1-r_j}{r_j}}{(\lvert\mathcal{E}\rvert-1)}$ to guarantee that $\bar{c}_i$ is within the range of $[0,1]$.

According to Equation~(\ref{eq:payback}), we set
\begin{equation} \label{eq:fpb}
g_i=
\begin{cases}
\displaystyle{\frac{\bar{c}_i}{\sum_{j\in\mathcal{E}}\bar{c}_j}}R-\displaystyle{\frac{1-r_i}{r_i}}\bar{c}_i, &\mbox{if $c_i\geqslant\bar{c}_i$},\\
\displaystyle{\frac{c_i}{\sum_{j\in\mathcal{E}}\bar{c}_j}}R-\displaystyle{\frac{1-r_i}{r_i}}c_i, &\mbox{if $c_i<\bar{c}_i$}.
\end{cases}
\end{equation}

Also, according to Equation~(\ref{eq:reputation}), $s$ updates all $r_i\in\mathcal{R_{\mathcal{E}}}$ as:
\begin{equation} \label{eq:fru}
r_i=
\begin{cases}
\alpha r'_i+(1-\alpha), &\mbox{if $c_i\geqslant\bar{c}_i$},\\
\displaystyle{\alpha r'_i+(1-\alpha)\frac{c_i}{\bar{c}_i}}, &\mbox{if $c_i<\bar{c}_i$}.
\end{cases}
\end{equation}

As demonstrated, CRI guarantees that (i) $s$ can recruit a proper number of the most credible applicants for sensing tasks, and (ii) recruited users can only obtain the maximum paybacks when they contribute no less than expected. Cheating behaviors will reduce their paybacks and opportunities of being recruited in future tasks.

\section{Evaluation}\label{sec:evaluation}

To validate the performance of CRI in real-world MCSs, we conducted extensive trace-driven simulations based on OMNeT++ 4.6, using real-world outdoor temperature data collected by taxis in Rome (hereinafter referred to as Rometrace) \cite{queensu}. In the following, we first present the simulation settings, and then show the evaluation results.

\subsection{Simulation Settings}\label{subsec:simulationsetting}

According to Rometrace, we constructed an MCS with a cloud server and 366 registered users. All users possessed outdoor temperature data opportunistically collected within $24$ hours. The server spontaneously announced temperature sensing tasks to the users. After receiving an announcement, a user who possessed data collected within $\pm60$ seconds autonomously applied for the task, and then uploaded corresponding report if it was recruited. The server provided paybacks and updated employee reputations based on CRI during the simulation.

For the parameter settings, we set the initial reputation $r_0=0.5$ and $\alpha=0.5$ in Equation~(\ref{eq:reputation}) for reasonable reputation bootstrapping and management. In addition, we set $\epsilon=0.1$ in Algorithm~\ref{alg:sensingtruthdiscovery} as the truth discovery convergence threshold. Again, according to Rometrace, each round of simulation lasts for $86400$ simulation seconds.

We collected the following four metrics to evaluate the impact of cheating behaviors\footnote{We considered user's cheating behavior in simulations as reporting a random observation within the range of $[2\celsius,24\celsius]$ (according to Rometrace).} on the MCS performance:

\begin{itemize}

\item \textbf{Discovered Truth (DT)} refers to the sensing truth discovered in a task, whose cumulative distribution reflects the sensing accuracy. Ideally, CRI should be able to prevent cheating behaviors from disrupting DT;

\item \textbf{Reputation (REP)} refers to the user reputation, whose cumulative distribution reflects the user's behavior in historical tasks. Ideally, CRI should be able to downgrade a cheater's REP in proportion to its cheating intensity;

\item \textbf{Payback (PB)\footnote{The payback of each task was normalized within the range of $[0,1]$ for effective comparisons.}} refers to what a user receives for accomplishing sensing tasks, which reflects the motivation of the user participating in future tasks. Ideally, CRI should be able to reduce the PB that a user can get if it cheats;

\item \textbf{Task Count (TC)} refers to the number of sensing tasks accomplished by a user, which reflects the popularity of the user. Ideally, CRI should be able to limit the probability of a cheater participating in MCS tasks.

\end{itemize}

For comparison, we ran a round of simulation without any cheating behavior as the baseline (\emph{i.e.} the no cheating scenario). Then, we analyzed the impact of cheating behaviors introduced by users with different properties. In following subsections, we depict the simulation results using either the Cumulative Distribution Figure (CDF) or the Time-Variance Figure (TVF) for a distinct demonstration.

\subsection{Impact of General Cheating Intensity}
\label{subsec:generalcheatingintensity}

\begin{figure}[!t]
\centering
\subfloat[Discovered Truth (CDF)]{\includegraphics[width=1.6in]{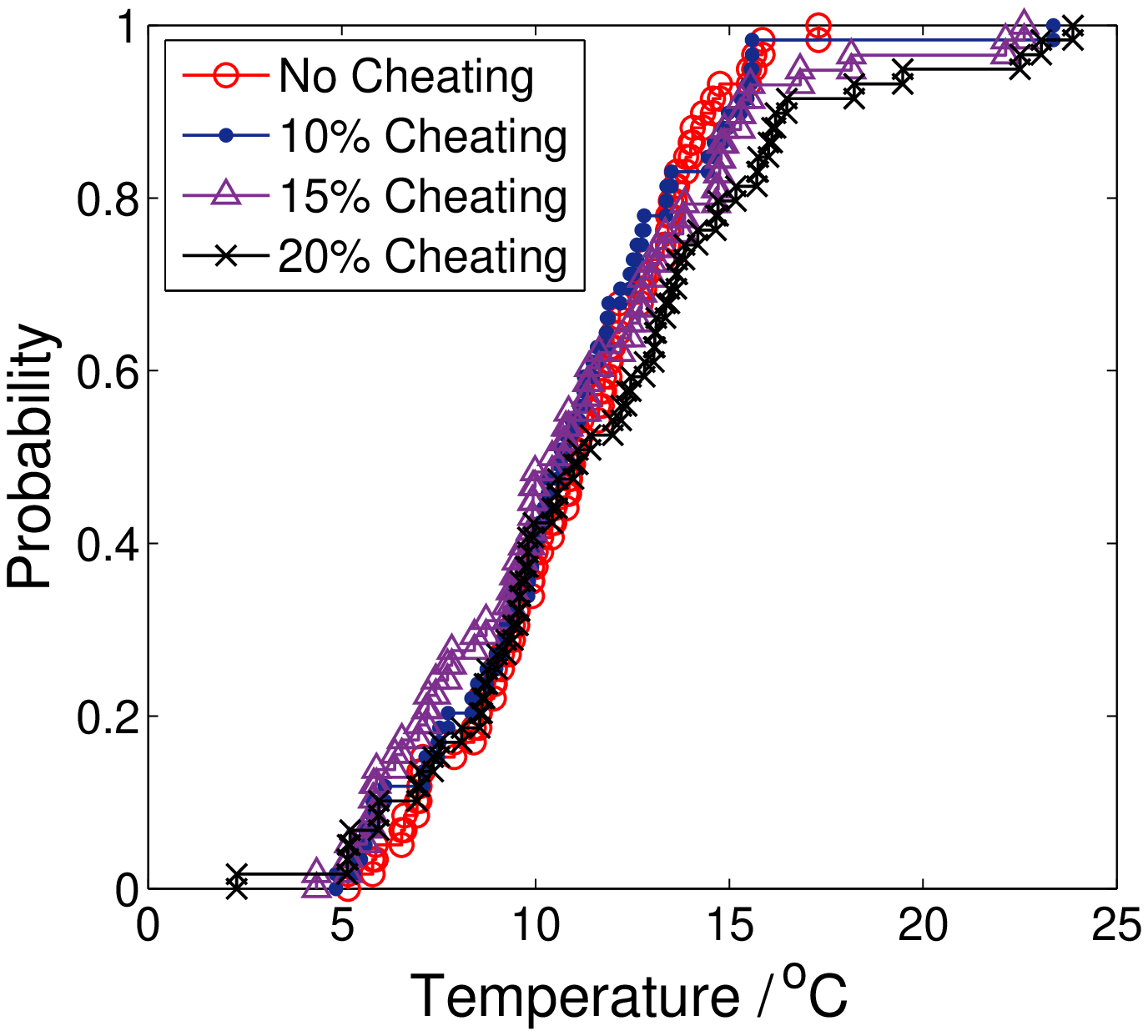}}
\vspace{-0.5em}
\hfil
\subfloat[Reputation (CDF)]{\includegraphics[width=1.6in]{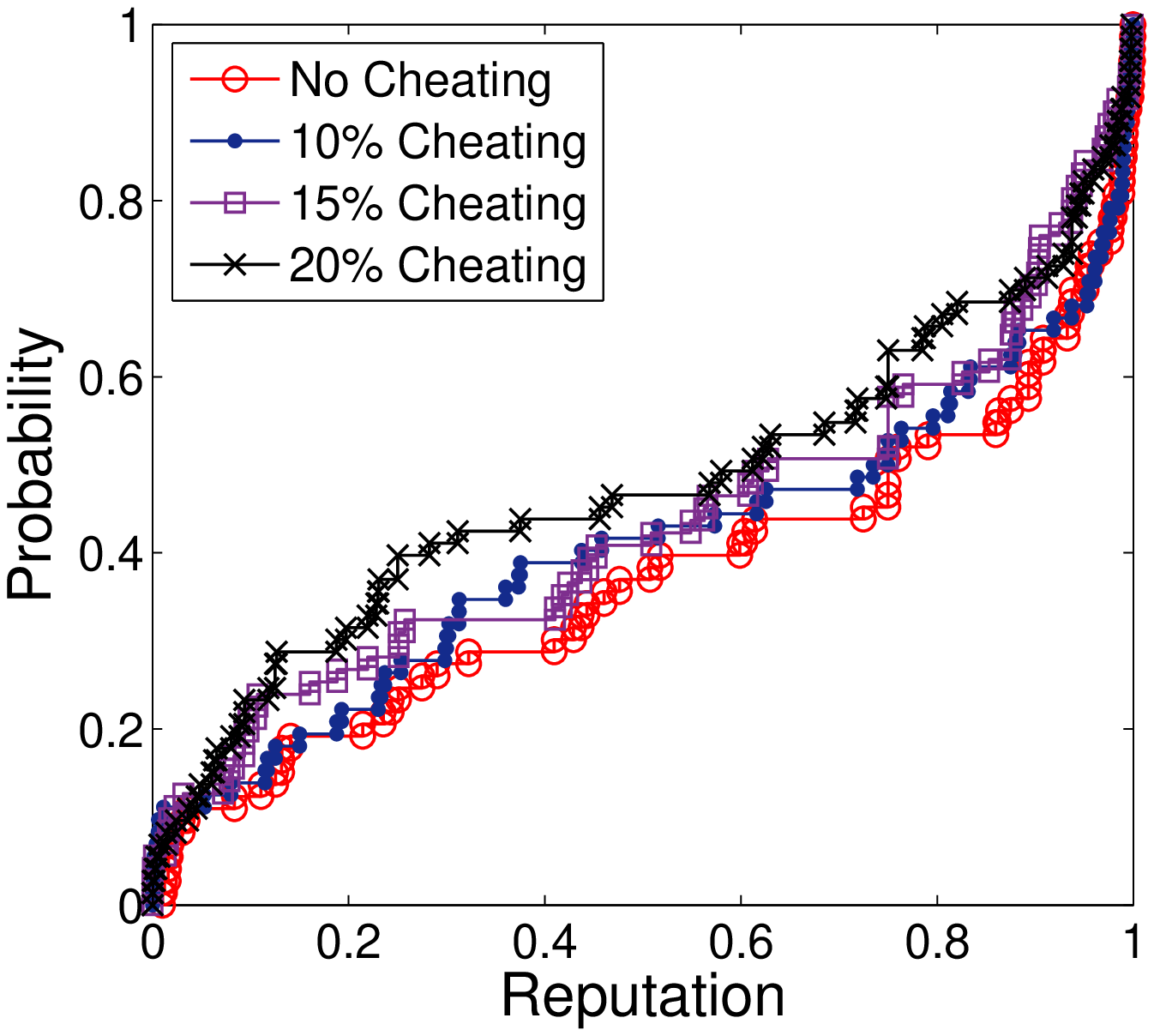}}
\hfil
\subfloat[Payback (CDF)]{\includegraphics[width=1.6in]{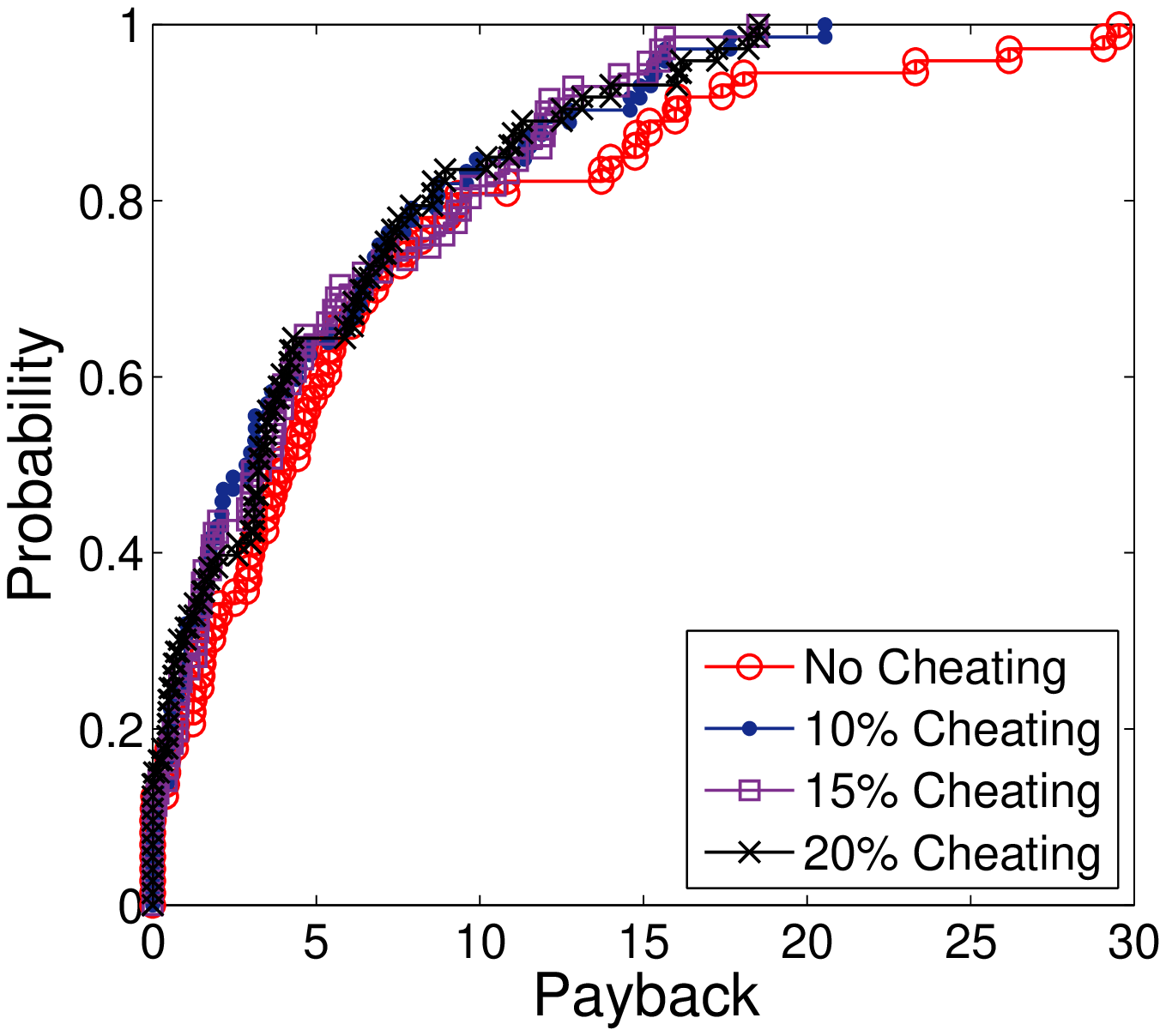}}
\hfil
\subfloat[Task Count (CDF)]{\includegraphics[width=1.6in]{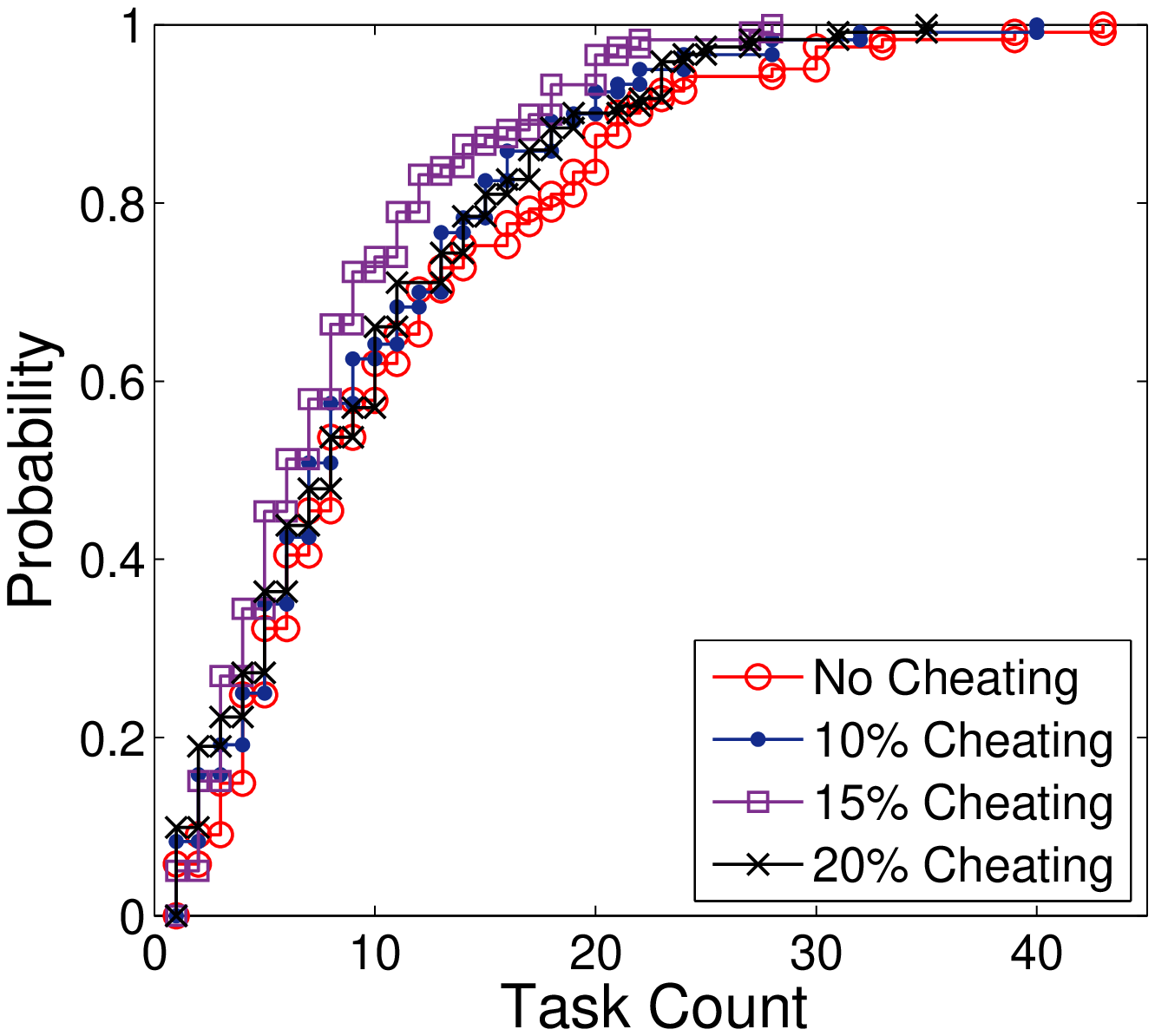}}
\caption{Impact of General Cheating Intensity}
\label{fig:ratio}
\end{figure}

In this set of simulations, to study the impact of general cheating behaviors with different intensities, we set all users in the MCS to introduce cheating behaviors with different probabilities (\emph{i.e.} 10\%, 15\% and 20\%)\footnote{The setting of these cheating intensities should be reasonable considering the well accepted fact that the MCS is a relatively good community with a limited ratio of malicious behaviors (\emph{e.g.} 4\% in \cite{Shen2015Enhancing}, or 10\% in \cite{Li2012Scalable}).} in all sensing tasks. The simulation result is illustrated in Figure~\ref{fig:ratio}.

\begin{figure}[!t]
\centering
\subfloat[Discovered Truth (CDF)]{\includegraphics[width=1.6in]{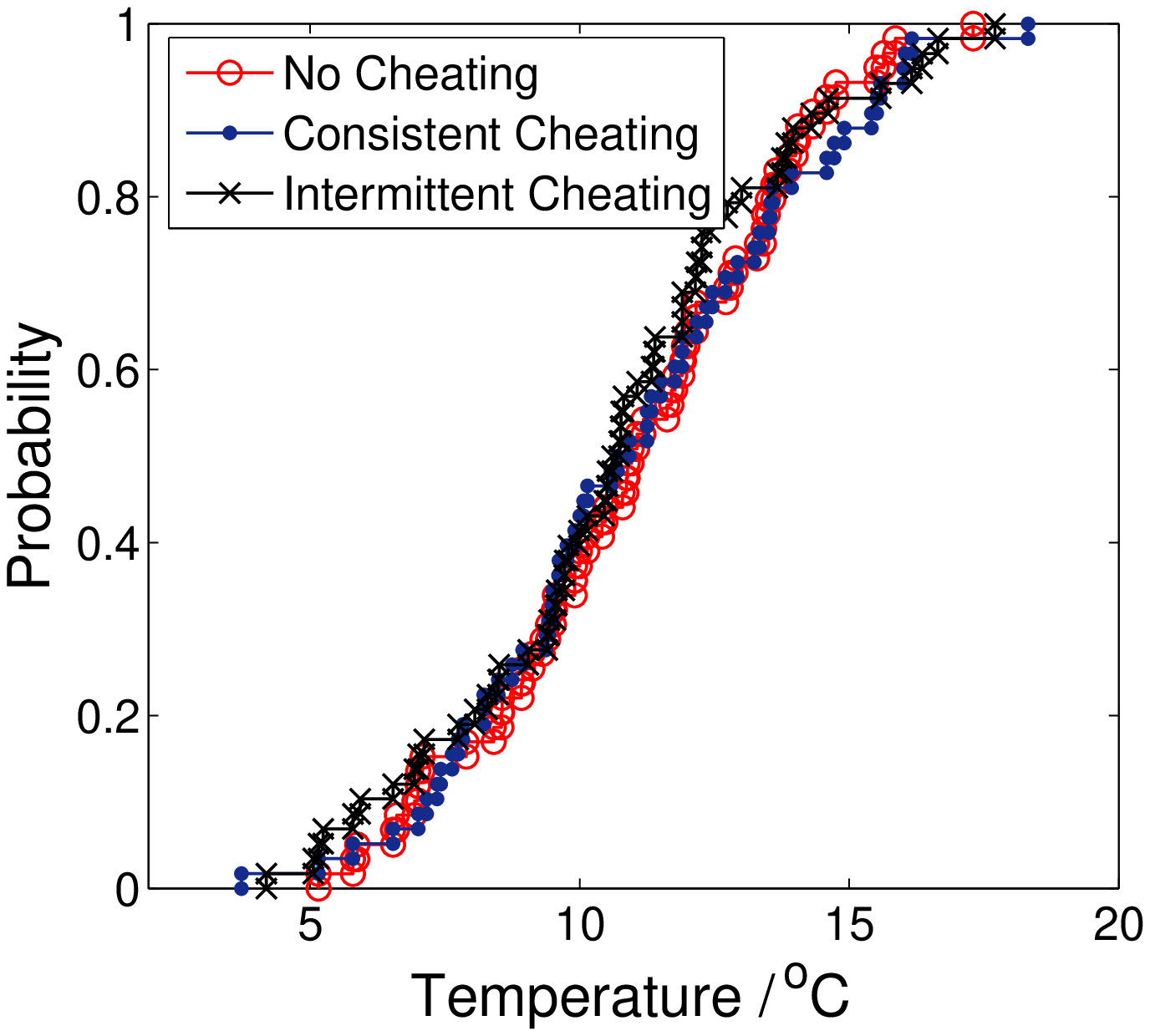}}
\vspace{-0.5em}
\hfil
\subfloat[Reputation (CDF)]{\includegraphics[width=1.6in]{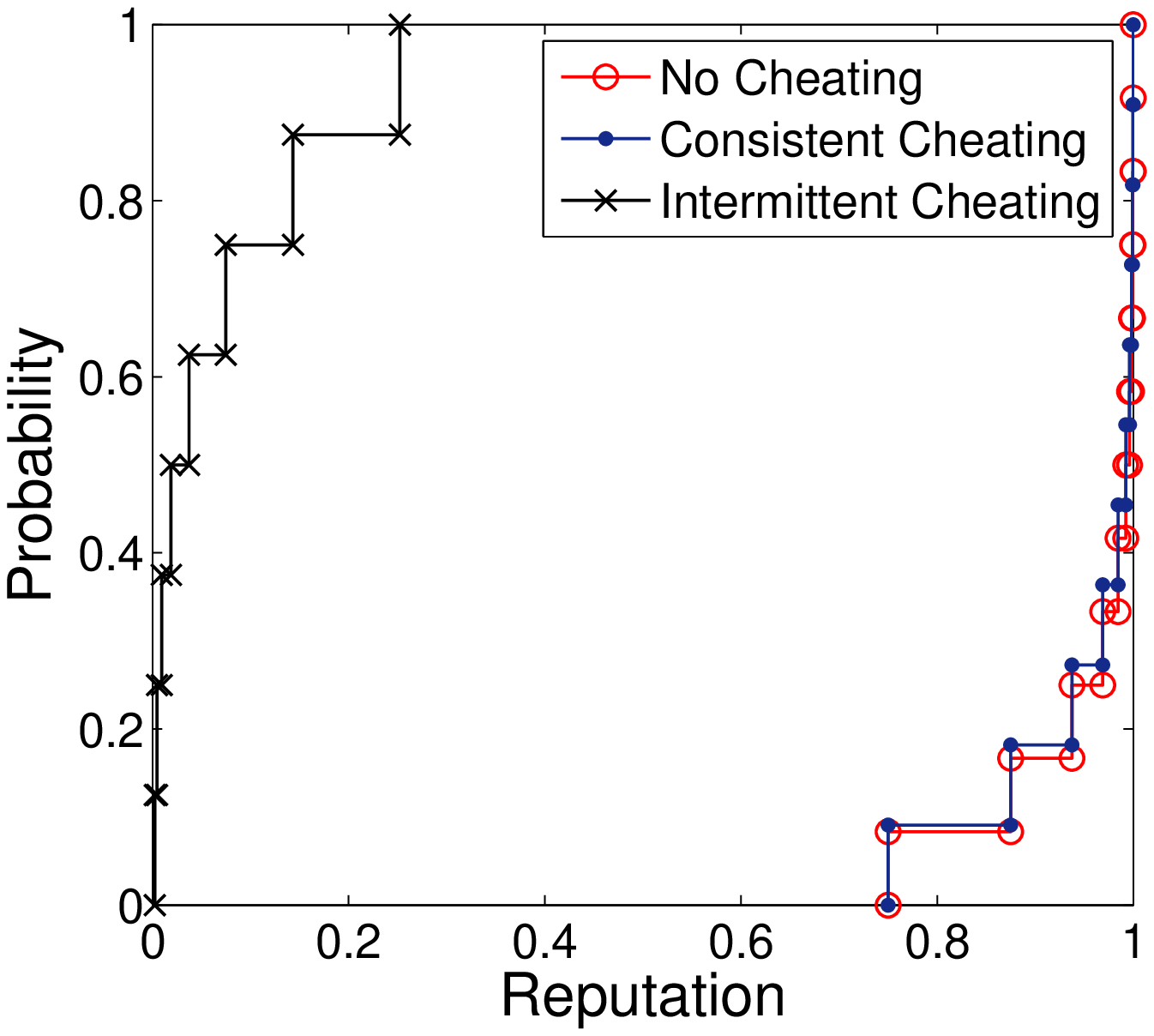}}
\hfil
\subfloat[Payback (TVF)]{\includegraphics[width=1.6in]{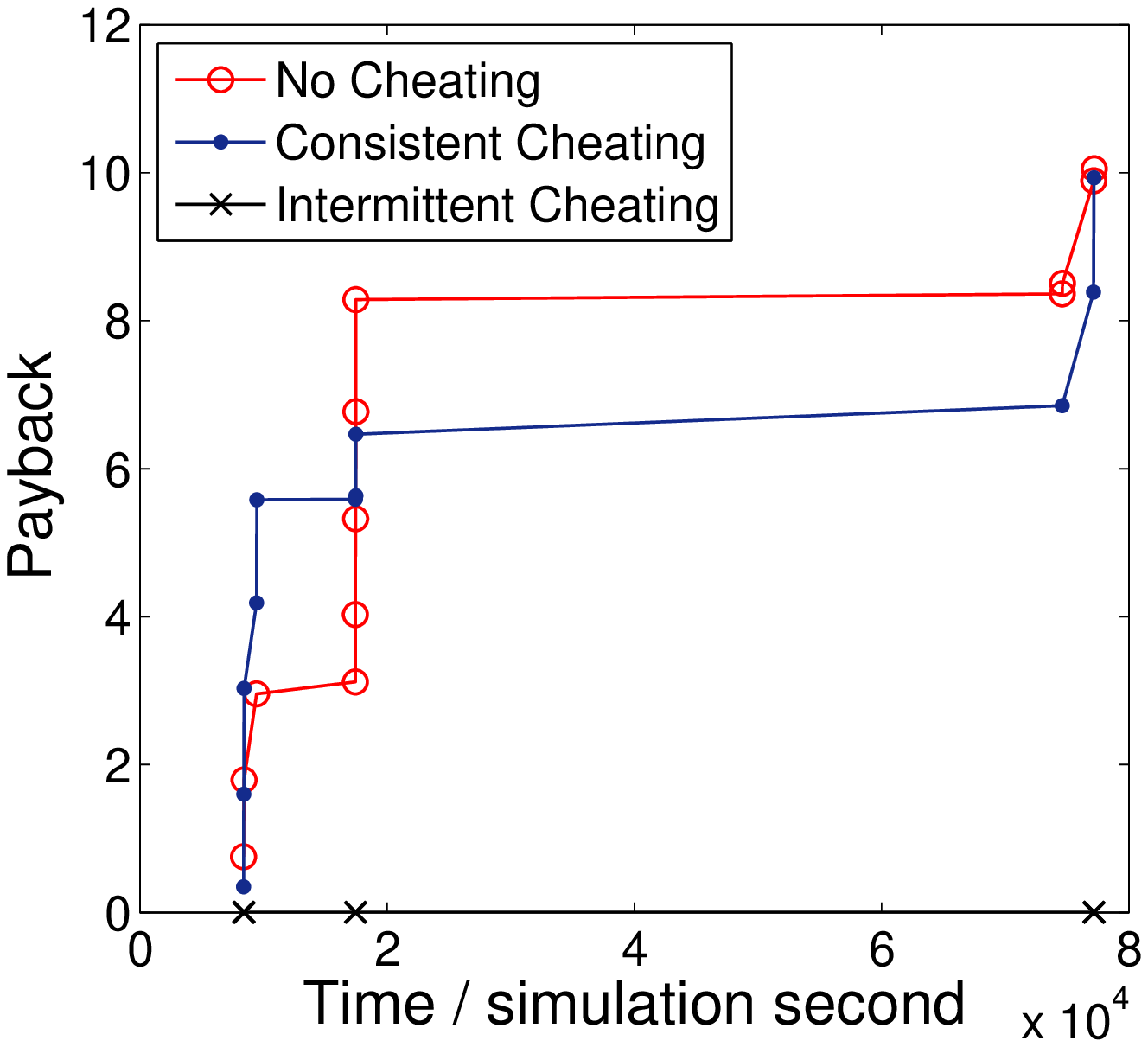}}
\hfil
\subfloat[Task Count (TVF)]{\includegraphics[width=1.6in]{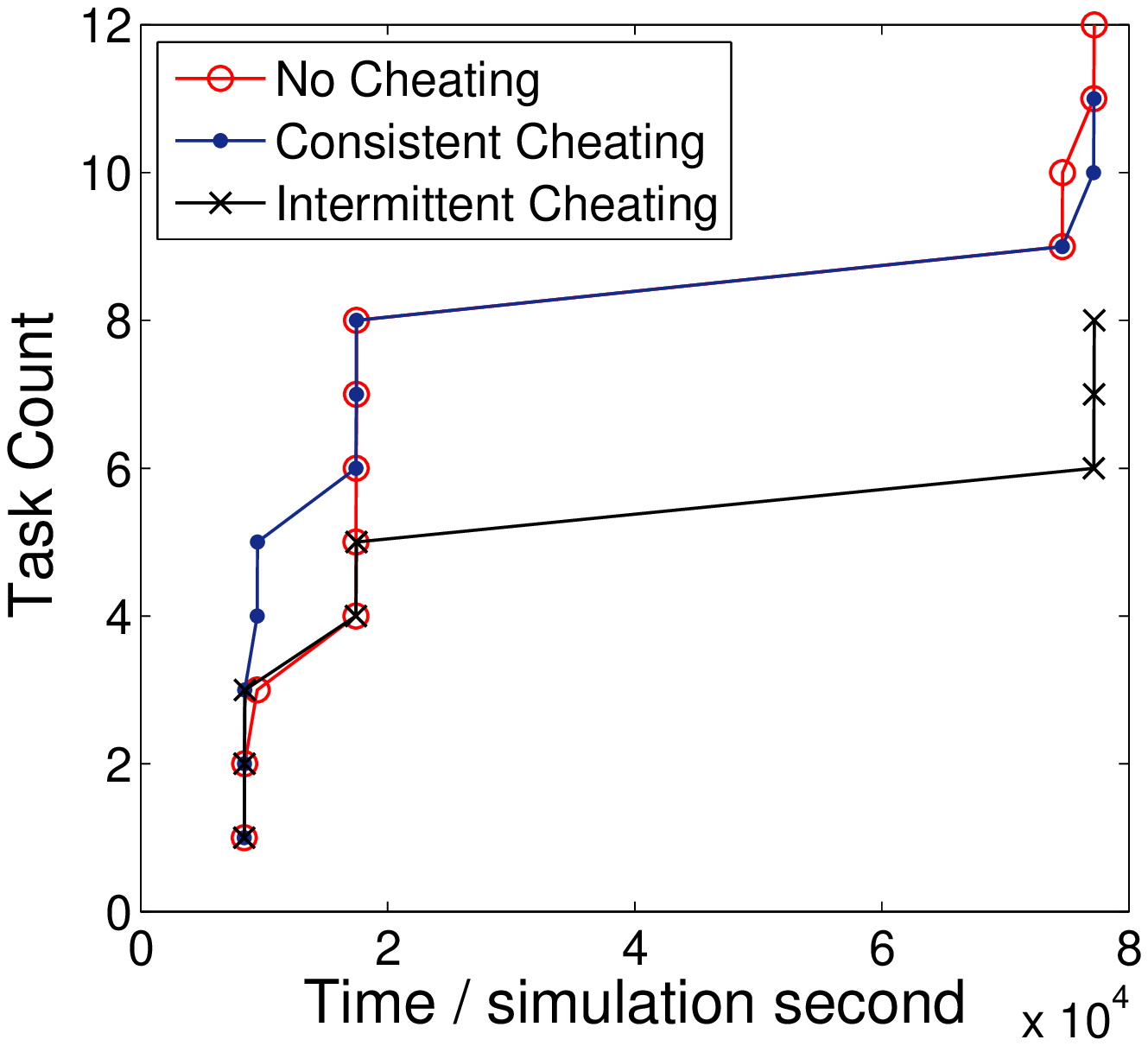}}
\caption{Impact of Cheating Behavior of TopR User}
\label{fig:topr}
\end{figure}

According to Figure~\ref{fig:ratio}(a), compared with the baseline scenario, the cumulative distribution of DT remains almost the same in all cheating scenarios. We can see that a disturbance on the average DT no more than $3.65\%$ (\emph{i.e.} 0.4\celsius) is introduced by general cheating behaviors with an intensity up to $20\%$. Such an impact is nearly negligible considering the practical temperature sensing requirement. CRI manages to effectively restrict the impact of general cheating behaviors on DT in both realistic and even harsher scenarios.

\begin{figure}[!t]
\centering
\subfloat[Discovered Truth (CDF)]{\includegraphics[width=1.6in]{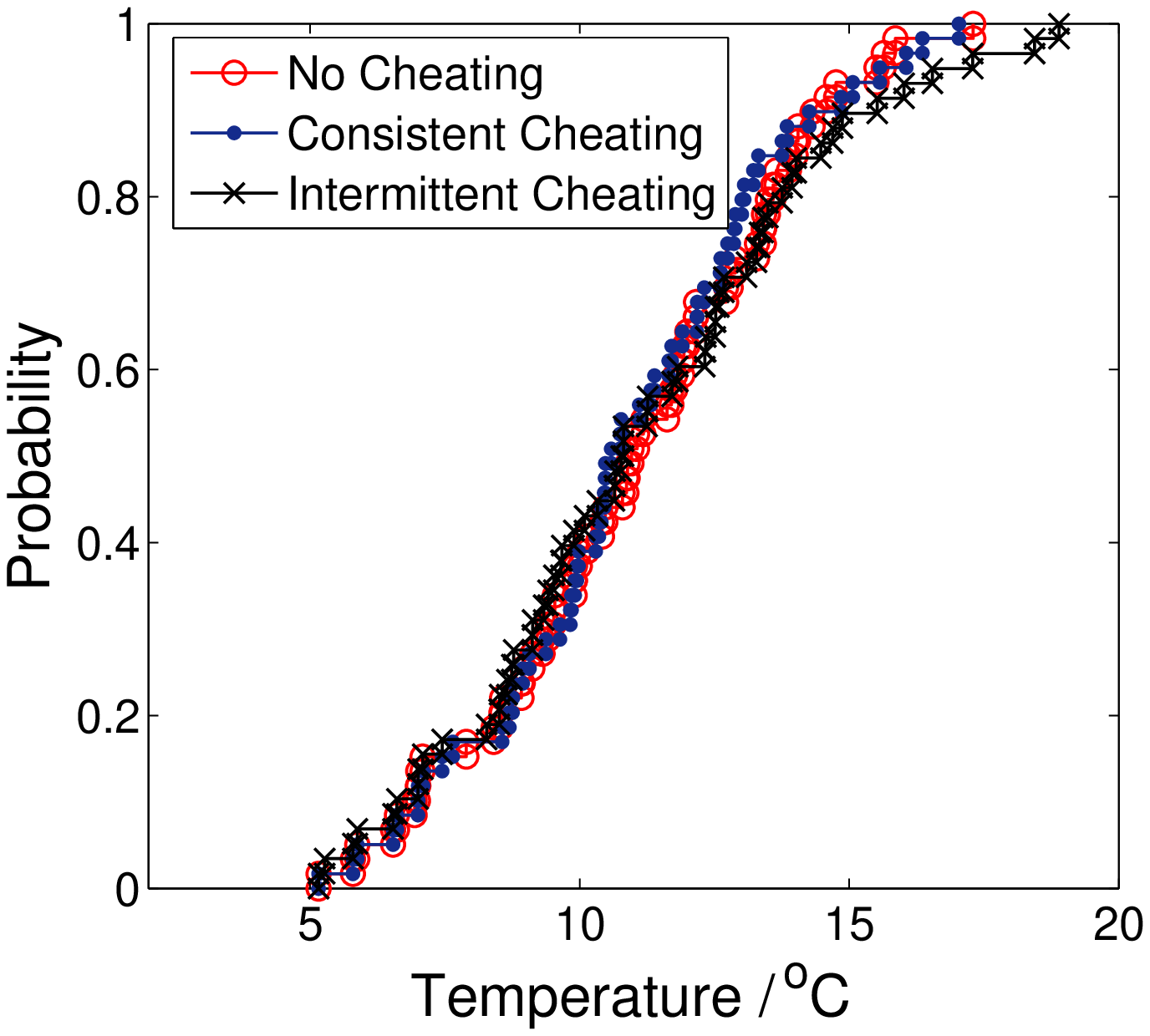}}
\vspace{-0.5em}
\hfil
\subfloat[Reputation (CDF)]{\includegraphics[width=1.6in]{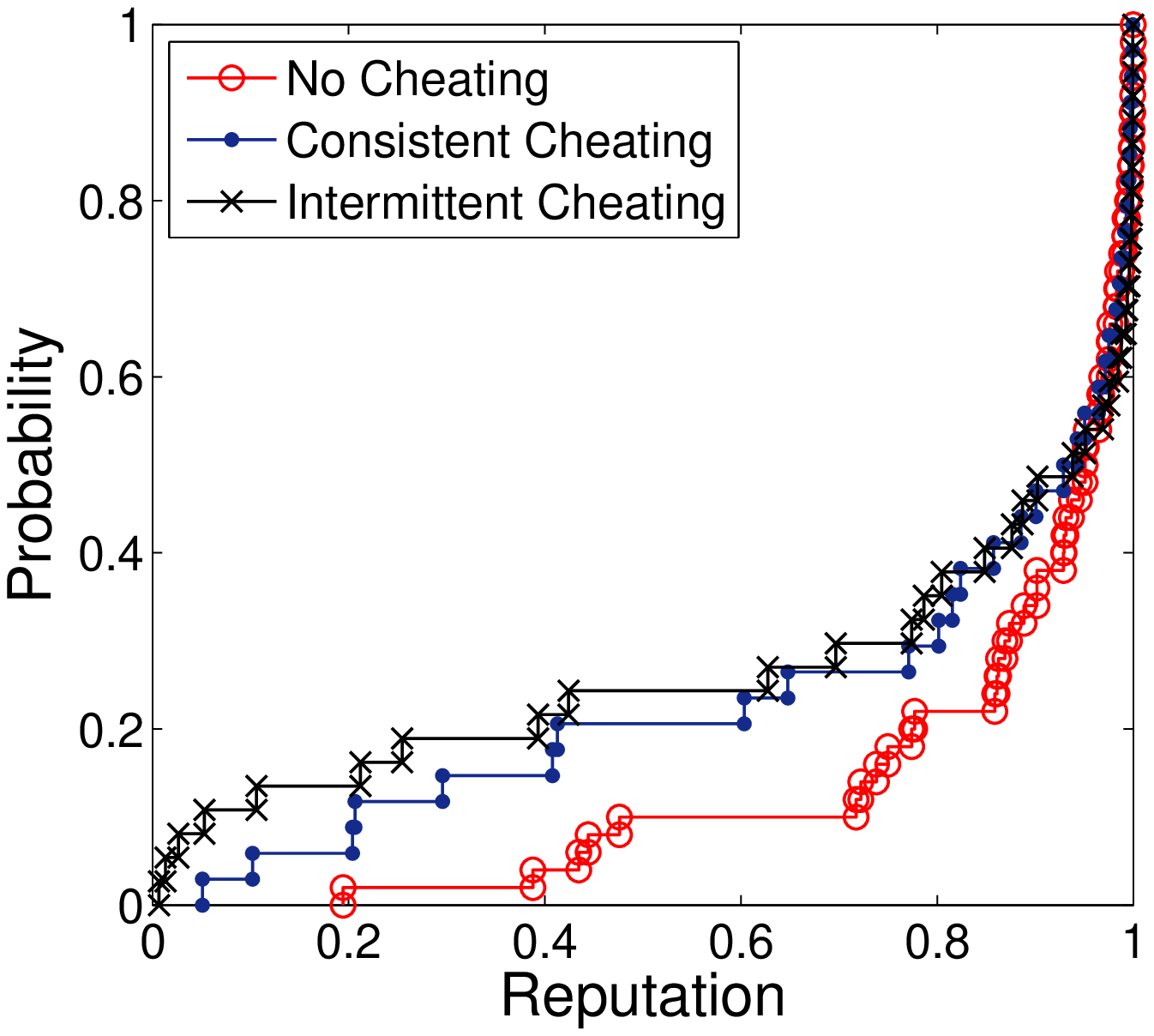}}
\hfil
\subfloat[Payback (TVF)]{\includegraphics[width=1.6in]{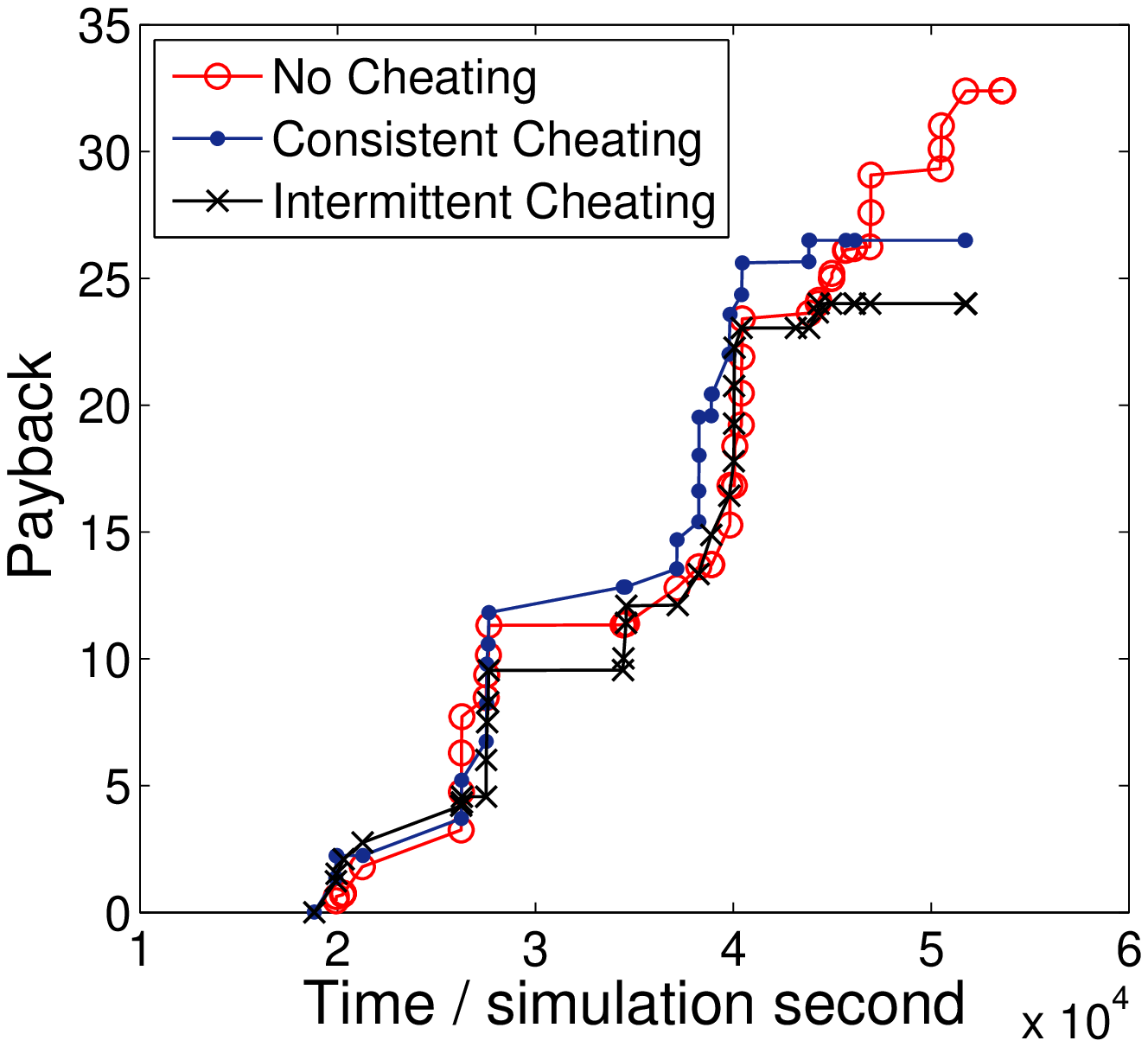}}
\hfil
\subfloat[Task Count (TVF)]{\includegraphics[width=1.6in]{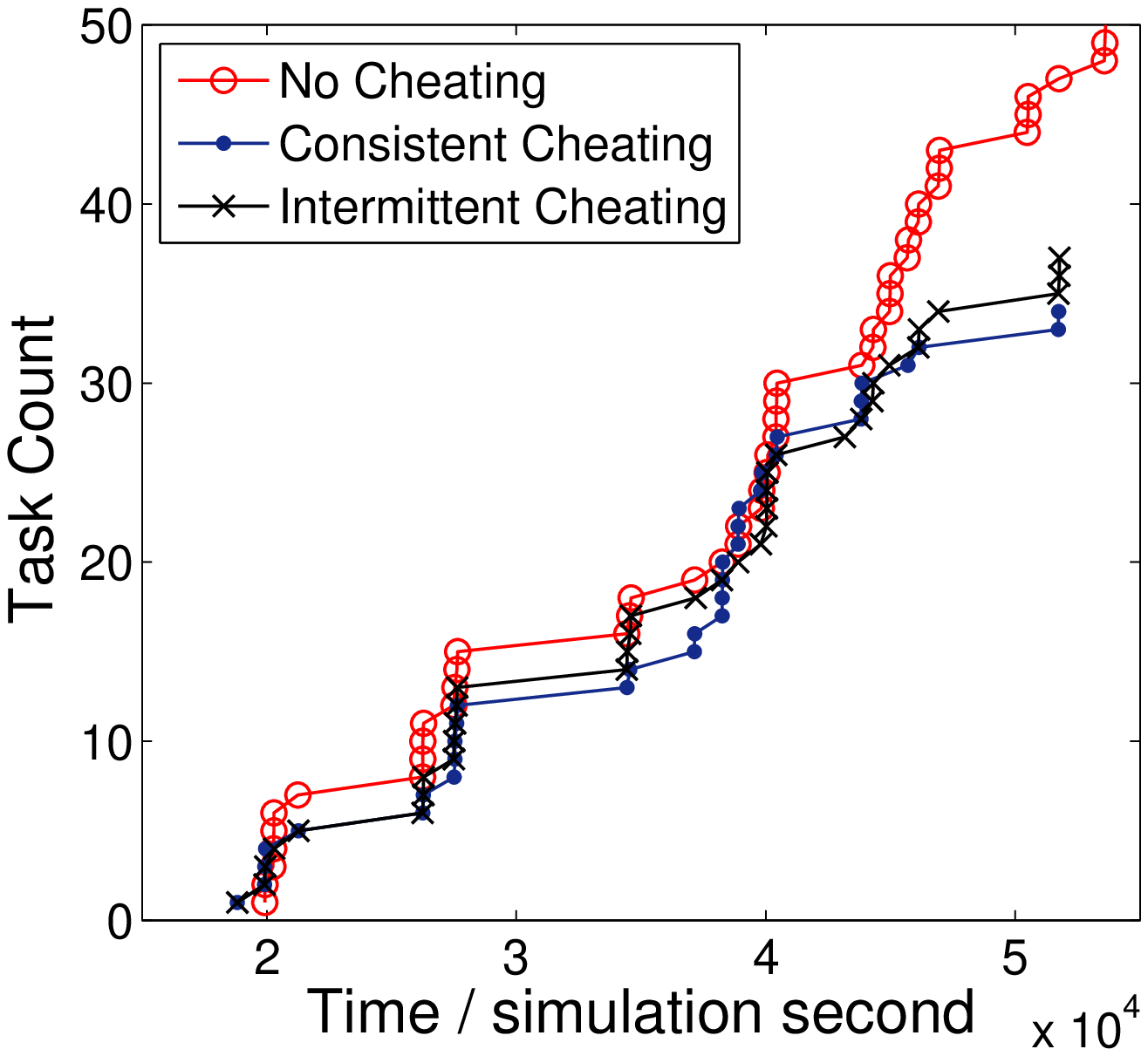}}
\caption{Impact of Cheating Behavior of TopP User}
\label{fig:topp}
\end{figure}

\begin{figure}[!t]
\centering
\subfloat[Discovered Truth (CDF)]{\includegraphics[width=1.6in]{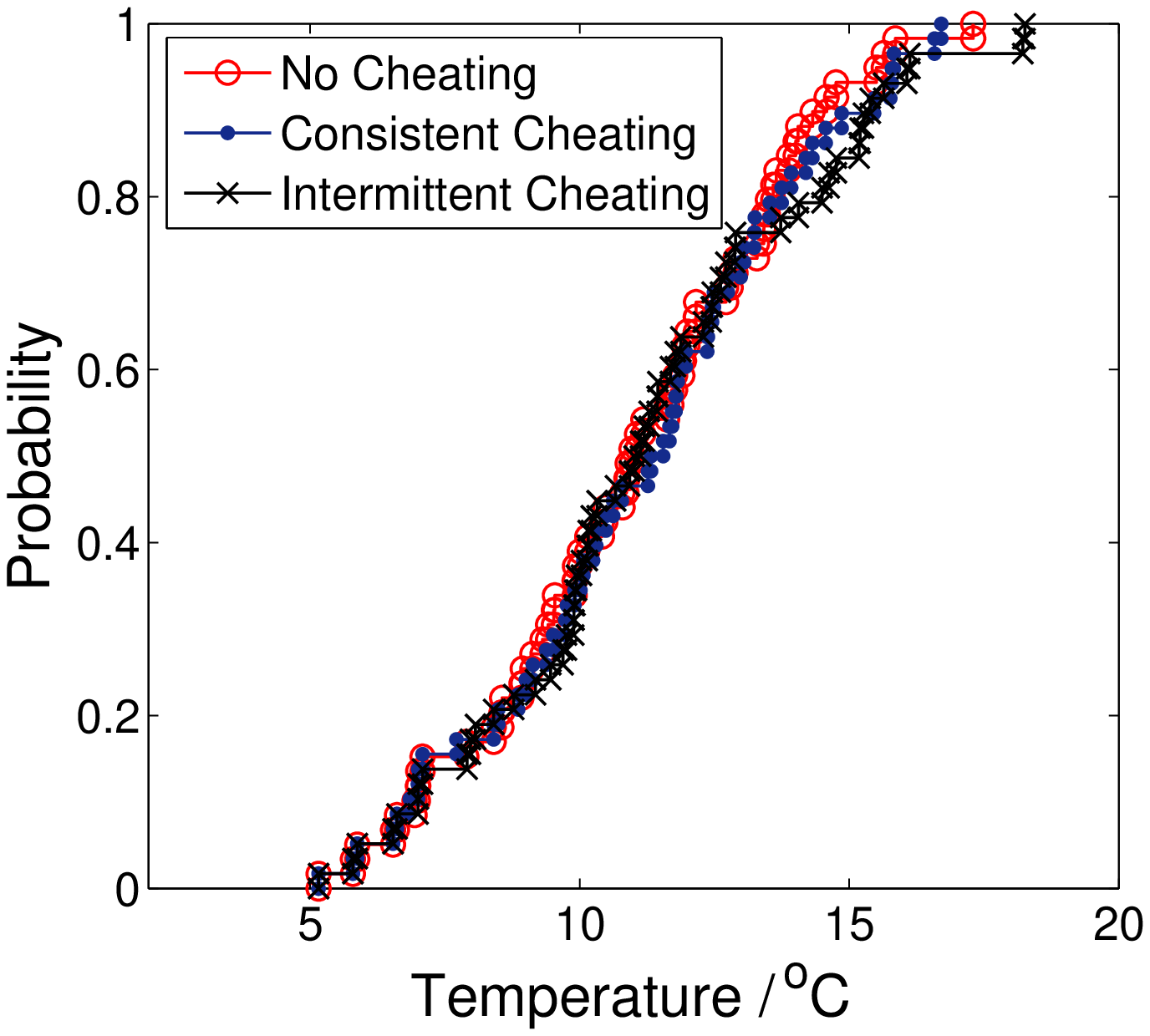}}
\vspace{-0.5em}
\hfil
\subfloat[Reputation (CDF)]{\includegraphics[width=1.6in]{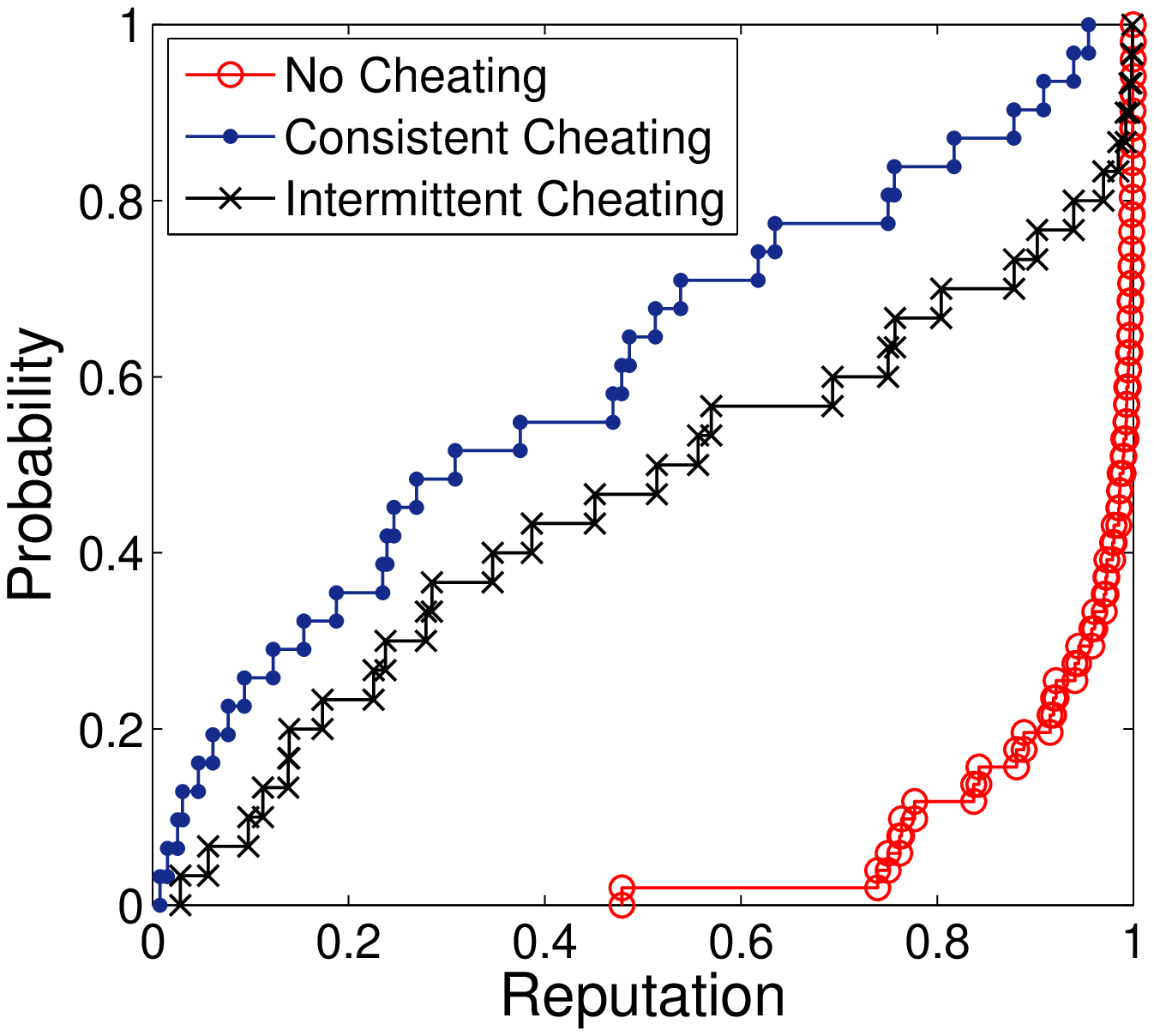}}
\hfil
\subfloat[Payback (TVF)]{\includegraphics[width=1.6in]{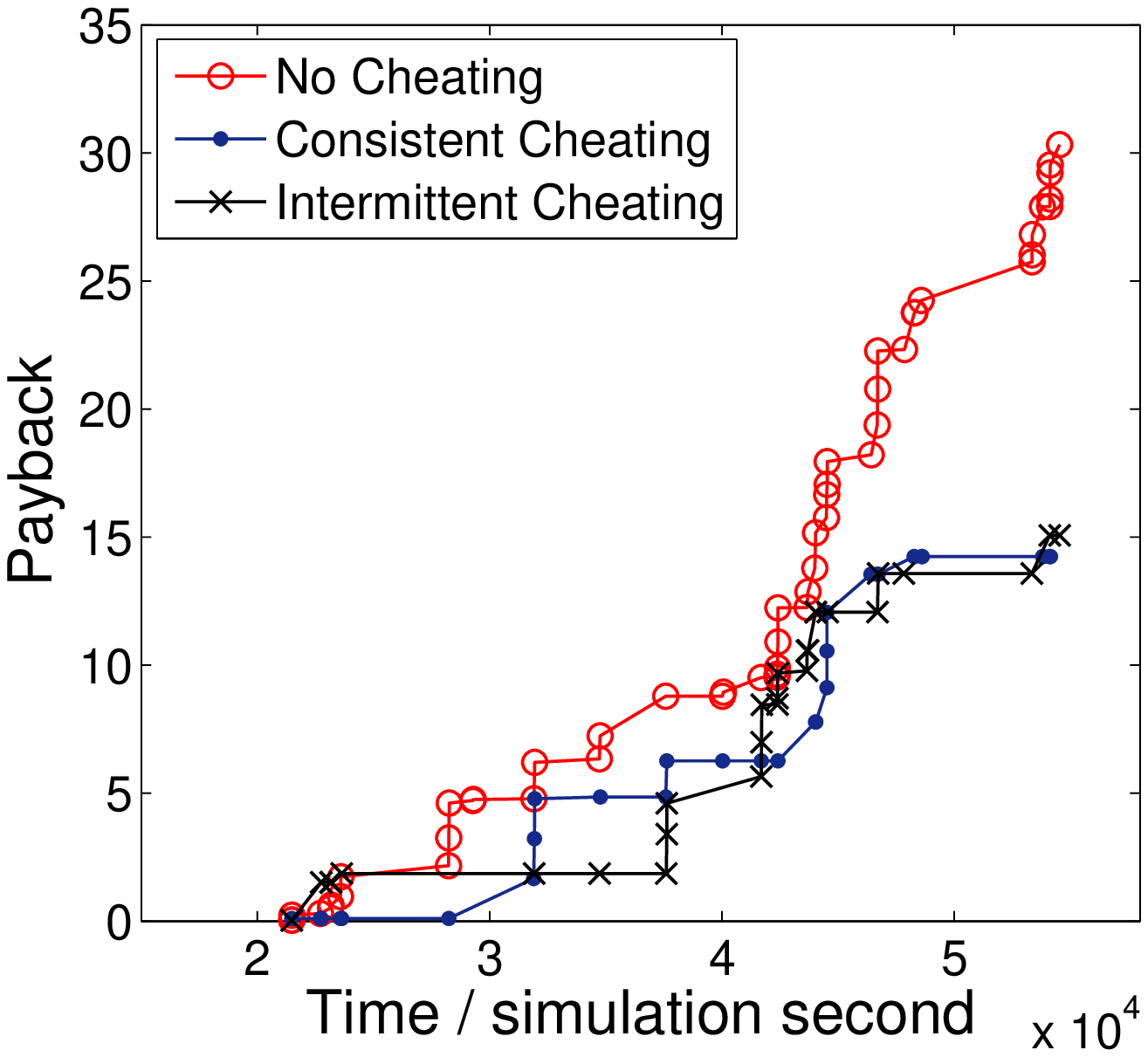}}
\hfil
\subfloat[Task Count (TVF)]{\includegraphics[width=1.6in]{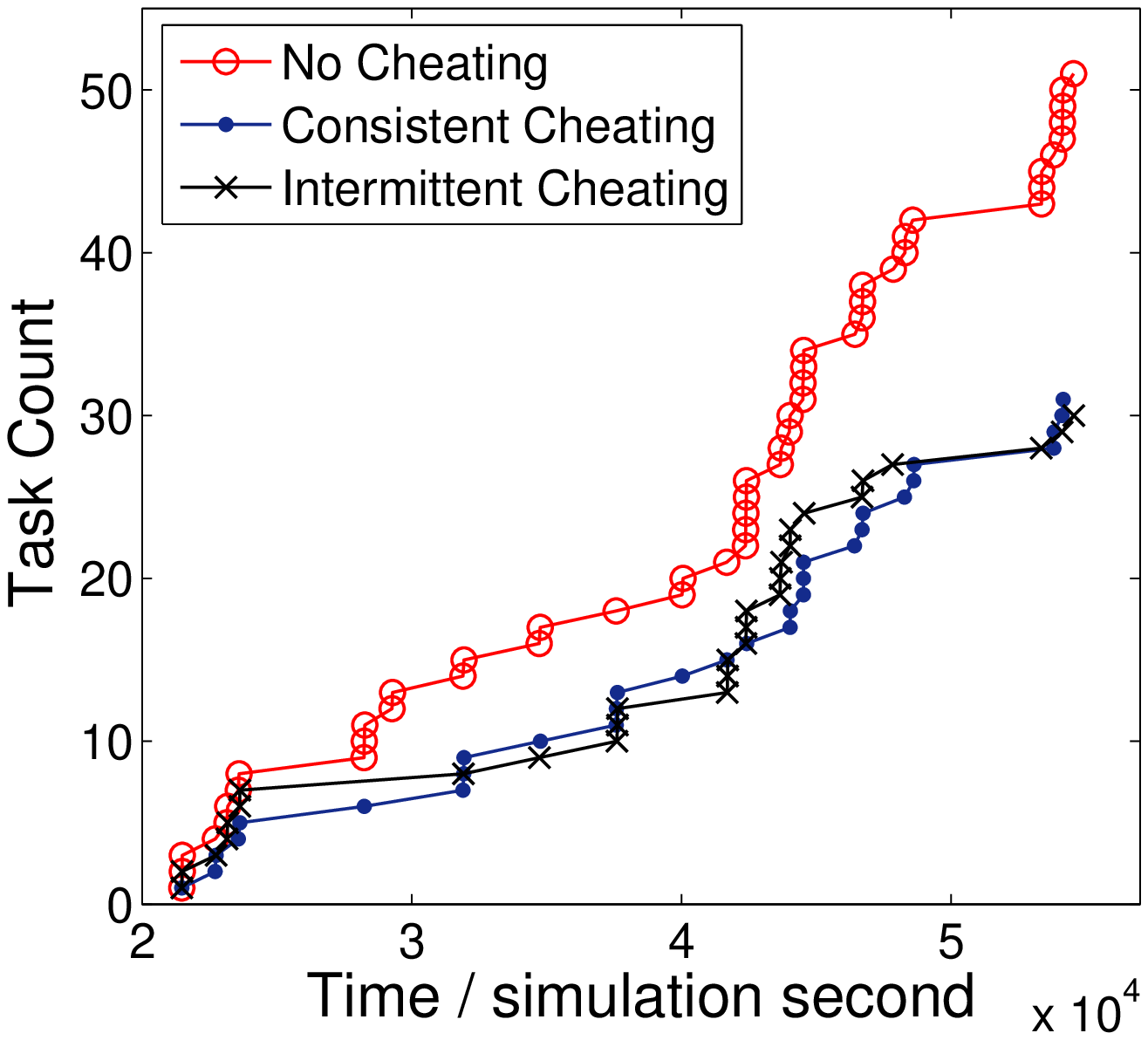}}
\caption{Impact of Cheating Behavior of TopC User}
\label{fig:topc}
\end{figure}

According to Figure~\ref{fig:ratio}(b), (c), and (d), when the cheating intensity increases, user's reputation, payback, and task count are correspondingly downgraded for at least $1.64\%$, $3.44\%$ and $2.20\%$, respectively. CRI manages to reduce the cheater's probability of being recruited in future tasks by reducing its reputation and payback autonomously, which will inherently restrict user's cheating intention.

\subsection{Impact of Cheaters with Different Properties} \label{subsec:impactofcheaterswithdifferentproperties}

In real-world MCSs, cheating behaviors of more trustworthy or active users may pose deeper impacts on the MCS's performance. In this set of simulations, depending on the simulation result of the baseline scenario, we separately set a user that (i) had the highest reputation (referred to as the TopR cheater), (ii) received the most paybacks (referred to as the TopP cheater), and (iii) accomplished the most tasks (referred to as the TopC cheater) to introduce cheating behaviors either consistently (\emph{i.e.} cheat with a 100\% probability) or intermittently (\emph{i.e.} cheat with a 50\% probability). The simulation results are illustrated in Figures~\ref{fig:topr}, \ref{fig:topp}, and~\ref{fig:topc}.

\textbf{TopR Cheater:} According to Figure~\ref{fig:topr}, neither the consistent nor the intermittent cheating of the TopR cheater caused obvious impact on DT (introduced disturbances of $0.18\%$ and $0.46\%$ on the average DT, respectively). Nonetheless, in comparison with the baseline scenario, REP of the TopR cheater was downgraded as long as there was cheating behavior ($1.04\%$ and $92.71\%$ lower, respectively). Correspondingly, both PB ($1.19\%$ and $99.98\%$ less, respectively) and TC ($8.33\%$ and $33.33\%$ less, respectively) of the TopR cheater decreased.

\textbf{TopP Cheater:} According to Figure~\ref{fig:topp}, neither consistent nor intermittent cheating behaviors of the TopP cheater caused obvious impact on DT (introduced disturbances of $0.73\%$ and $0.82\%$ on the average DT, respectively). Meanwhile, REP of the TopP cheater ($10.34\%$ and $14.94\%$ lower, respectively) was significantly downgraded. Similarly, it's PB ($18.18\%$ and $25.93\%$ less, respectively) and TC ($32\%$ and $26\%$ less, respectively) declined dramatically because of cheating.

\textbf{TopC Cheater:} According to Figure~\ref{fig:topc}, DT was obviously affected by neither consistent nor intermittent cheating behaviors of the TopC cheater (introduced disturbances of $0.91\%$ and $0.37\%$ on the average DT, respectively). In turn, it's REP ($58.51\%$ and $42.55\%$ lower, respectively) was severely downgraded in cheating scenarios. Also, PB ($53.05\%$ and $50.35\%$ less, respectively) and TC ($39.22\%$ and $41.18\%$ less, respectively) decreased significantly.

According to the results above, it is well demonstrated that CRI manages to encourage users to report honestly (with their best efforts) in sensing tasks for higher payback, reputation, and recruiting opportunities in practical MCSs.

\section{Conclusion} \label{sec:conclusion}

In this paper, we developed CRI for MCSs to guarantee crowdsensing accuracy by encouraging mobile users to provide quality data without cheating for the maximum paybacks. To be specific, CRI enables MCS server to autonomously recruit as many as credible users as task employees, and employees will obtain the maximum payback only if they contribute honestly as their reputations indicate. Via theoretical analysis, we demonstrated the feasibility and correctness of our design. To evaluate the performance of CRI in practical MCSs, we conducted extensive simulations based on real-world crowdsensing data. The results show that CRI manages to guarantee the sensing accuracy under realistic cheating intensities (up to 20\% of total reports). Meanwhile, cheating behaviors from users with selected advantages (\emph{i.e.} higher reputations, more received paybacks, and more accomplished tasks) can be effectively resisted as well. Our future work is to develop a privacy-preserving CRI, which encourages honest user behaviors without jeopardizing sensitive user information including identities, locations, and living patterns.

\end{document}